\documentclass[12pt]{article}
\usepackage{amssymb,amsmath,epsfig}
\usepackage{graphicx}
\usepackage{amsfonts,graphicx,amssymb,amsmath,amsthm,epsfig}
\usepackage{float}
\allowdisplaybreaks
\begin{document}

\title{\textbf{Analyzing Gravastar Structure with the Finch-Skea Metric
in Extended Modified Symmetric Teleparallel Gravity}}
\author{Iqra Ibrar \thanks{iqraibrar26@gmail.com}~ and
M. Sharif \thanks {msharif.math@pu.edu.pk}\\
Department of Mathematics and Statistics, The University of Lahore,\\
1-KM Defence Road Lahore-54000, Pakistan.}
\date{}
\maketitle
\begin{abstract}
This study analyzes the physical features of a gravastar within the
$f(\mathcal{Q}, \mathbb{T})$ gravity framework, where $\mathcal{Q}$
is the non-metricity scalar and $\mathbb{T}$ is the trace of the
energy-momentum tensor. Gravastars present a viable alternative to
black holes, featuring a central de Sitter core, a surrounding thin
shell and a dynamic layer in the Schwarzschild exterior that
separates these two regions. Using the Finch-Skea metric, the
necessary field equations for the core and shell are derived, while
the Israel junction conditions maintain a seamless connection
between the inner and outer regions. This work extensively explores
crucial aspects such as energy distribution, proper length, energy
conditions, entropy and the equation of state parameter. The model's
stability is studied through the effective potential, redshift,
causality conditions and adiabatic index. Our results highlight the
essential role of modified gravity in maintaining the structural
viability and stability of gravastars.
\end{abstract}
\textbf{Keywords}: Israel formalism; $f(\mathcal{Q},\mathbb{T})$
gravity; Gravastars.\\
\textbf{PACS}:04.70.Bw; 04.50.kd; 04.40.Dg.

\section{Introduction}

Astrophysical research reveals that the universe commenced its
expansion approximately 13.8 billion years ago, leading to a rapid
cooling of its temperature. This vigorous expansion facilitated the
formation of large-scale cosmic structures, including stars, planets
and galaxies, each exhibiting unique and remarkable features.
Various theoretical models have been developed to explain the
mechanisms behind the creation of these celestial bodies. Einstein
formulation of General Relativity (GR) fundamentally transformed our
understanding of the cosmos by linking the motion of massive objects
to the curvature of spacetime induced by gravitational fields.
Following this, Edwin Hubble \cite{a} observations demonstrated that
galaxies are moving away from us at accelerating speeds, a
phenomenon strongly supported by empirical data. The driving force
behind this accelerated expansion is attributed to an elusive entity
known as dark energy (DE), characterized by its significant negative
pressure. Despite its success in describing cosmic acceleration, the
standard cold dark matter ($\Lambda$CDM) model, which incorporates
the cosmological constant $(\Lambda)$ from GR, faces critical
challenges. Notably, there exists a stark discrepancy of about 120
orders of magnitude between the observed value of $\Lambda$ and the
theoretically predicted vacuum energy density. Additionally, the
model struggles to explain why the current value of $\Lambda$ is
comparable to the present matter density, raising questions about
its completeness in describing the universe evolution \cite{b}.

The prevailing big bang theory offers a comprehensive framework for
understanding the universe origin and its subsequent expansion from
an initial singularity (a point of infinite density and
temperature). However, in the earliest moments of the universe, the
singularity existed in an extremely energetic regime where GR
becomes ineffective due to expected quantum phenomena. This
limitation has prompted the scientific community to explore
alternative models to address these cosmological conundrums. One
such innovative approach is the bounce theory, which proposes a
cyclic model of cosmic expansion and contraction, thereby avoiding
the initial singularity inherent in the big bang scenario. By
positing recurrent phases of expansion and contraction, bounce
theory provides a robust foundation for developing models that
circumvent the singularity problem. Moreover, challenges like the
flatness problem, monopole abundance and the horizon dilemma further
indicate that the standard GR based cosmological model is
insufficient for explaining the universe under extreme conditions.
Consequently, there is a growing necessity for theories that
integrate quantum mechanics with gravity to achieve a unified
description of spacetime and gravitational interactions, paving the
way for a more complete understanding of the universe fundamental
nature.

In the quest to extend and refine our understanding of gravitational
phenomena beyond the framework of GR, various modified theories of
gravity have been proposed. Among these, symmetric teleparallel
gravity (STG) presents a distinctive paradigm within the spectrum of
gravitational theories, fundamentally diverging from GR by
attributing gravitational interactions to the non-metricity of
spacetime rather than its curvature \cite{c}. In STG, the affine
connection is meticulously chosen to be both torsion-free and
curvature-free, thereby isolating non-metricity as the sole
geometric entity responsible for mediating gravitational phenomena.
This sophisticated framework leverages metric-affine geometry,
enabling a more adaptable and versatile description of spacetime
dynamics. By focusing solely on non-metricity, STG enables the
formulation of modified gravitational models, such as
$f(\mathcal{Q})$ gravity \cite{d}, which encompass crucial geometric
properties for gravitational interactions. A lot of significant work
has been done in this modified framework \cite{17aa}-\cite{17hh}.
Alternative theories and observational constraints has been examined
in \cite{19aa}-\cite{19gg}.

Building on the foundational principles of STG, the
$f(\mathcal{Q},\mathbb{T})$ gravity theory marks a notable
progression in the field of modified gravitational frameworks
\cite{e}. In this extended model, the $\mathcal{Q}$ continues to
characterize the geometric attributes of spacetime arising from the
connection inability to maintain metric compatibility, while the
trace of the energy-momentum tensor (EMT) introduces a direct
coupling between matter and the geometric structure. By generalizing
the traditional Einstein-Hilbert action to incorporate an arbitrary
function $f(\mathcal{Q},\mathbb{T})$, this theory allows for a more
intricate interaction between the gravitational field and matter
distribution, potentially addressing persistent cosmological and
astrophysical challenges such as DE and the universe accelerated
expansion. The integration of both non-metricity and the trace of
EMT in $f(\mathcal{Q},\mathbb{T})$ gravity provides novel insights
into the dynamics of cosmic structures and the evolution of the
universe. Furthermore, this theoretical framework opens new avenues
for explaining phenomena that remain unresolved within GR, offering
additional degrees of freedom through the function
$f(\mathcal{Q},\mathbb{T})$. Consequently, this theory not only
deepens our understanding of gravitational interactions but also
presents a versatile platform for exploring alternative models that
could unify various aspects of cosmological observations and
theoretical predictions, thereby enriching our comprehension of
gravity role in the fundamental constituents of the universe. This
has become subject of great interest in scientific community due to
its crucial implications in the field of cosmology and astrophysics
\cite{18aa}-\cite{18kk}.

This theory has drawn the attention of researchers because of its
significant theoretical consequences and its vital function in
clarifying various astronomical events. Arora et al. \cite{f}
examined a modified gravity model to investigate cosmic
acceleration, utilizing constraints from Hubble and supernova data.
Their results indicate that this model effectively tackles the DE
issue and necessitates further research in cosmology. Bhattacharjee
\cite{g} investigated gravitational baryogenesis in this gravity,
finding that it can significantly enhance baryon to entropy ratio.
Godani and Samanta \cite{h} explored $f(\mathcal{Q},\mathbb{T})$
gravity through a non-linear model, deriving cosmological
implications consistent with supernova data and the $\Lambda$CDM
model. Agrawal et al. \cite{i} examined an extension of STG
concerning late-time cosmic acceleration. Their work derives
dynamical parameters and validates non-singular matter bounce models
through energy conditions and stability analysis. Shiravand et al.
\cite{j} investigated cosmological inflation within
$f(\mathcal{Q},\mathbb{T})$ gravity, deriving modified slow-roll
parameters and spectral indices. Their findings demonstrate
alignment with Planck 2018 observational data by appropriately
constraining free parameters. Tayde et al. \cite{k} analyzed the
potential for wormholes within the framework of
$f(\mathcal{Q},\mathbb{T})$ gravity. Recently, we have presented the
ghost, generalized ghost and generalized ghost pilgrim DE models
within a cohesive theoretical framework \cite{l}.

The study of self-gravitating celestial bodies has garnered
significant attention from researchers due to their remarkable
properties and their impact on both cosmology and astrophysics.
Accurate theoretical models are essential for understanding these
entities, particularly in the context of stellar gravitational
collapse, which results in the creation of dense, compact
structures. Although such formations exhibit intriguing geometric
properties, they also present a fundamental issue due to the
existence of a singularity surrounded by an event horizon. To
address this, Mazur and Mottola \cite{m} proposed the gravastar as
an alternative to black holes, offering a solution that eliminates
the singularity issue. Gravastars feature a core of vacuum energy
with negative pressure, likely associated with DE, which prevents
the formation of a singularity and produces a repulsive force.

Unlike black holes, gravastars do not have an event horizon and are
characterized by an outer shell composed of normal matter. Although
no direct observations of gravastars have been made, their unique
characteristics, such as gravitational lensing and specific shadow
patterns \cite{o1}, could eventually lead to their detection.
Gravastars also provide an alternative explanation for the
gravitational waves detected by LIGO, fueling continued research
interest \cite{o2}. Building on Mazur and Mottola model \cite{n},
Visser and Wiltshire \cite{o} simplified the gravastar structure,
reducing it to three layers: a de Sitter core, a Schwarzschild
exterior and a thin-shell of stiff fluid in between. This refined
model removes both the event horizon and de Sitter horizon, offering
a non-singular alternative to black holes \cite{p}. Furthermore,
gravastars may contain exotic forms of matter like DE, allowing
researchers to explore high energy physics and the behavior of
matter under extreme gravitational conditions \cite{q}.

Gravastars have increasingly captured the interest of the scientific
community keen to explore their unique structural characteristics.
Visser and Wiltshire \cite{r} investigated the impact of radial
perturbations on gravastar stability, successfully identifying
configurations that remain stable under such disturbances. Carter
\cite{19} discovered new stable spherically symmetric gravastar
solutions, emphasizing different parameter ranges and their
qualitative behaviors of the equation of state (EoS) parameter.
Bilic et al. \cite{s} analyzed the geometry of gravastars with
substantial radii and masses, providing deeper insights into their
extensive structural properties. Ghosh et al. \cite{t} examined
gravastar geometries, enabling a thorough analysis of their various
characteristics. Das et al. \cite{23} analyzed the properties of the
intermediate shell of gravastars within the modified theory of
gravity and illustrated some of these features using graphical
methods. Shamir and Ahmad \cite{23aa} investigated the structure of
gravastars in $f(\mathcal{G},\mathbb{T})$ gravity, where
$\mathcal{G}$ represents the Gauss-Bonnet term and analyzed various
physical properties, suggesting that gravastars might offer stable,
non-singular solutions without the presence of an event horizon.
Sharif and Waseem \cite{25} studied gravastars within the same
framework using conformal motion and found that these objects do not
possess singularities. Ghosh and his colleagues \cite{22}
investigated gravastar configurations by applying the Karmarkar
condition to both the interior and shell layers. Bhatti et al.
\cite{22b} investigated a singularity-free gravastar model within
the framework of $f(\mathcal{R}, \mathcal{G})$ gravity, where
$\mathcal{R}$ is Ricci scalar and discussed its physical properties.
Pradhan et al. \cite{11} proposed gravastars as alternatives to
black holes, explaining the universe accelerated expansion. This
study presents a stable, singularity-free gravastar model,
emphasizing its interior, thin-shell properties and outer vacuum
region based on Schwarzschild geometry. Mohanty and Sahoo \cite{11a}
studied a gravastar in $f(\mathcal{Q})$ gravity using the
Krori-Barua metric, focusing on field equations, physical properties
and potential differences across the thin-shell.

Recent studies in various gravitational theories have significantly
contributed to the understanding of gravastars, providing valuable
insights into their physical characteristics and potential detection
methods. Javed et al. \cite{13a} explored how electric charge
impacts gravastars in this gravity. Mohanty et al. \cite{14a}
studied charged gravastars, focusing on energy density, entropy and
the EoS. They also discussed how future radio telescopes might
detect the shadow of a gravastar, helping to differentiate it from a
black hole event horizon. Odintsov and Oikonomou \cite{15a} studied
static neutron stars in the context of various inflationary models,
comparing them with theoretical and observational constraints. They
found that all models are compatible with constraints, with maximum
masses in the mass gap region and explored the potential to
distinguish inflationary attractors. Teruel et al. \cite{17a}
reported the first existence of gravastar configurations in
$f(\mathcal{R}, \mathbb{T})$ gravity by developing the field
equations and studying four models. Sharif et al. \cite{18a}
presented a novel gravastar solution in non-conservative Rastall
gravity, deriving singularity-free radial metric potentials and
investigating various properties of the gravastar shell, concluding
it as a viable alternative to black holes.

To address the limitations of the Dourah and Ray metric \cite{37} in
modeling compact astrophysical objects, researchers introduced the
Finch-Skea metric as a more effective alternative. This metric is
simple in mathematics and aligns with physical reality, making it
easier to understand compact stars like neutron stars. The main
reason for using the Finch-Skea metric is that it meets the
necessary energy conditions, ensuring that the models being created
are realistic. Its analytical tractability, meaning it can be solved
easily, helps in accurately solving the field equations without
difficulties of numerical simulations. Additionally, the Finch-Skea
metric has shown compatibility with observational data, especially
in reproducing the mass-radius relationships observed in neutron
stars. This alignment with empirical findings increases the
predictive power of the models, bridging the gap between theoretical
ideas and astronomical observations. Furthermore, the Finch-Skea
metric is a valuable tool for exploring extensions in modified
theories of gravity. By adapting this metric within alternative
frameworks, we can investigate possible deviations and assess the
robustness of stellar models under different theoretical
assumptions.

Several studies have successfully applied the Finch-Skea metric in
various astrophysical contexts, further supporting its relevance and
motivating its use in our analysis. Finch and Skea \cite{38} refined
this metric to better align with relativistic stellar models. Bhar
\cite{45} proposed an anisotropic strange star model with the
Chaplygin EoS, based on the Finch-Skea ansatz, matching
observational data of SAX J 1808.4-3658 and free from central
singularity. Paul and Dey \cite{44} analyzed compact stars using the
Finch-Skea metric, finding isotropy in 4D but anisotropy in higher
dimensions. Their study examines physical parameters and the
feasibility of stellar models in varying spacetime dimensions.
Banerjee et al. \cite{a1} investigated compact astrophysical objects
governed by the DE EoS using the Finch-Skea metric. Their solutions
are consistent with the observed maximum mass limits of compact
stars, address stability concerns and provide insights into exotic
astrophysical phenomena. Sharif and Naz \cite{a2} studied gravastar
structures in $f(\mathcal{R}, \mathbb{T}^{2})$gravity using the
Finch-Skea metric, finding singularity-free solutions with smooth
matching to Schwarzschild spacetime and viable as black hole
alternatives. Dayanandan et al. \cite{a3} developed a deformed
Finch-Skea anisotropic solution using gravitational decoupling and
tested its physical viability for neutron stars and white dwarfs,
highlighting anisotropy role in preventing gravitational collapse.
Mustafa et al. \cite{a4} studied anisotropic compact stars in
$f(\mathcal{Q})$ gravity using the Finch-Skea metric, analyzing
stability, mass-radius relations and compactness for PSR J0437-4715,
finding that compactness increases with density. Shahzad et al.
\cite{a5} derived a new solution in Rastall theory with a
quintessence field using the Finch-Skea ansatz, validating it for
five compact stars through physical and graphical analysis,
fulfilling criteria for a viable stellar model. Rej et al. \cite{a6}
proposed a DE stellar model using the Finch-Skea ansatz and
vanishing complexity condition, demonstrating stability and
compatibility with observed compact objects, predicting masses
beyond the GR limit. Karmakar et al. \cite{a7} proposed a polytropic
star model in 5D Einstein-Gauss-Bonnet gravity using the Finch-Skea
ansatz, analyzing EXO 1785-248, and confirming the model realism
through physical viability and the effects of the Gauss-Bonnet
coupling constant. Das et al. \cite{a8} proposed a relativistic
model of anisotropic compact stars coupled with DE using the
Finch-Skea metric, analyzing PSR J0348+0432 and demonstrating that
the stiffness of the EoS varies with the DE coupling parameter.
These studies underscore the versatility and robustness of the
Finch-Skea metric in modeling various types of compact astrophysical
objects and highlight its potential for further exploration in
modified gravity theories, especially for understanding gravastar
stability and structure.

In this paper, we aim to explore the geometric and physical aspects
of a gravastar, focusing on its structure and behavior within the
context of modified gravity theory. This study investigates the
geometry of gravastars using the Finch-Skea metric within the
framework of $f(\mathcal{Q}, \mathbb{T})$ gravity. The paper is
organized as follows. Section \textbf{2} introduces the formulation
of $f(\mathcal{Q}, \mathbb{T})$ gravity, followed by the derivation
of field equations for a spherically symmetric spacetime using the
Finch-Skea metric. Section \textbf{3} explores the configuration of
gravastars. In section \textbf{4}, we provide the necessary junction
conditions for smoothly connecting the inner and outer regions of
the gravastar. Section \textbf{5} describes several important
properties of the model in the context of $f(\mathcal{Q},
\mathbb{T})$ gravity. Finally, the entire analysis is summarized in
the concluding section.

\section{Basic Concepts of $f(\mathcal{Q}, \mathbb{T})$ Gravity}

The following discussion outlines the core principles of the
$f(\mathcal{Q}, \mathbb{T})$ theory. In GR, the Levi-Civita
connection serves as the foundation \cite{1-a}, being both
torsion-free and compatible with the metric. This connection is
derived from the metric itself and its first derivatives. By easing
this constraint, we can define two separate rank-3 tensors: one
linked to the antisymmetric component of
$\hat{\Gamma}^{\alpha}_{[\gamma\eta]}$ and the other related to the
covariant derivative of the metric tensor
\begin{equation}\label{1}
T_{\gamma\eta}^{\alpha}=2\hat{\Gamma}^{\alpha}_{[\gamma\eta]},~~
\nabla_{\sigma}g_{\gamma\eta}\neq 0=\mathcal{Q}_{\gamma\eta\sigma},
\end{equation}
the most comprehensive connection, encompassing all possible
contributions, is represented as follows
\begin{equation}\label{2}
\hat{\Gamma}^{\varphi}_{\gamma\eta}=\mathcal{L}^{\varphi}_{~\gamma\eta} +
\mathcal{C}^{\varphi}_{~\gamma\eta}+\Gamma^{\varphi}_{~\gamma\eta}.
\end{equation}
Within this framework, $\mathcal{L}^{\varphi}_{~\gamma\eta}$ denotes
the disformation tensor, $\mathcal{C}^{\varphi}_{~\gamma\eta}$
signifies the contortion tensor and the Levi-Civita connection is
indicated by $\Gamma^{\varphi}_{\gamma\eta}$, which can be expressed
as follows \cite{2}
\begin{eqnarray}\label{3}
\mathcal{L}^{\varphi}_{~\gamma\eta}&=&\frac{1}{2} g^{\varphi\sigma}
(\mathcal{Q}_{\gamma\eta\sigma}+\mathcal{Q}_{\eta\gamma\sigma}
-\mathcal{Q}_{\varphi\gamma\eta}),
\\\label{5}
\mathcal{C}^{\varphi}_{~\gamma\eta}&=&\hat{\Gamma}^{\varphi}_{[\gamma\eta]}+
g^{\varphi\sigma}g_{\gamma\vartheta}\hat{\Gamma}^{\vartheta}_{[\eta\sigma]}+
g^{\varphi\sigma}g_{\eta\vartheta}\hat{\Gamma}^{\vartheta}_{[\gamma\sigma]},
\\\label{4}
\Gamma^{\varphi}_{\gamma\eta}&=&\frac{1}{2}g^{\varphi\sigma}
(g_{\sigma\eta,\gamma}+g_{\sigma\gamma,\eta}-g_{\gamma\eta,\sigma}).
\end{eqnarray} Consequently, the $\mathcal{Q}$ tensor can also be
derived as
\begin{equation}\label{6}
\mathcal{Q}_{\varphi\gamma\eta}=-g_{\gamma\eta,\varphi}+
g_{\eta\sigma}\hat{\Gamma}^{\sigma}_{\gamma\varphi}+g_{\sigma\gamma}
\hat{\Gamma}^{\sigma}_{\eta\varphi}.
\end{equation}
In the absence of torsion, Weyl-Cartan geometry simplifies to Weyl
geometry, where the connection is determined by the metric and
non-metricity. In this particular situation, the contortion tensor
disappears, meaning the connection is solely determined by the
non-metricity tensor \cite{3}. For symmetric connections, the
Levi-Civita connection can be articulated in terms of the
disformation tensor as follows
\begin{equation}\label{7}
\Gamma^{\varphi}_{~\gamma\eta}=-\mathcal{L}^{\varphi}_{~\gamma\eta}.
\end{equation}

Gravitational effects are described in the following non-covariant
formulation \cite{4}
\begin{equation}\label{8}
\mathcal{S}=\frac {1}{16\pi}\int
g^{\gamma\eta}(\Gamma^{\varsigma}_{\sigma\gamma}\Gamma^{\sigma}
_{\eta\varsigma}-\Gamma^{\varsigma}_{\sigma\varsigma}\Gamma^{\sigma}
_{\gamma\eta})\sqrt{-g} d^{4}x.
\end{equation}
Applying Eq.\eqref{7}, the action is represented as follows
\begin{equation}\label{9}
\mathcal{S}=-\frac {1}{16\pi}\int
g^{\gamma\eta}(\mathcal{L}^{\varsigma}_{\sigma\gamma}\mathcal{L}^{\sigma}
_{\eta\varsigma}-\mathcal{L}^{\varsigma}_{\sigma\varsigma}\mathcal{L}^{\sigma}
_{\gamma\eta})\sqrt{-g} d^{4}x.
\end{equation}
Another representation of this action is
\begin{equation}\label{10}
\mathcal{S}=\frac {1}{16\pi}\int\mathcal{Q}\sqrt{-g} d^{4}x,
\end{equation}
where
\begin{equation}\label{11}
\mathcal{Q}=-g^{\gamma\eta}(\mathcal{L}^{\varsigma}_{\sigma\gamma}\mathcal{L}^{\sigma}
_{\eta\varsigma}-\mathcal{L}^{\varsigma}_{\sigma\varsigma}\mathcal{L}^{\sigma}
_{\gamma\eta}).
\end{equation}
Lets now extend STG by introducing an arbitrary function of
$\mathcal{Q}$. In this extended framework, the Einstein-Hilbert
action is modified as follows
\begin{equation}\label{13}
\mathcal{S}=\frac {1}{16\pi}\int (\sqrt{-g}f(\mathcal{Q})+L_{m})
d^{4}x,
\end{equation}
where, $L_{m}$ represents the matter Lagrangian, which results in
the $f(\mathcal{Q})$ theory. This theory leads to the
$f(\mathcal{Q}, \mathbb{T})$ framework when combined with the trace
of the EMT. The action is expressed as
\begin{equation}\label{14}
\mathcal{S}=\frac{1}{16\pi}\int \sqrt{-g}f(\mathcal{Q},
\mathbb{T})d^{4}x+\int \sqrt{-g}L_{m}d^{4}x.
\end{equation}
The following expressions provide the traces of the non-metricity
tensor
\begin{equation}\label{15}
\mathcal{Q}_{\gamma}=\mathcal{Q}_{\gamma~~\varphi}^{~~\varphi} ,\quad
\mathcal{\tilde{Q}_{\gamma}}= \mathcal{Q}^{\varphi}_{~~\gamma\varphi}.
\end{equation}
The superpotential related to $\mathcal{Q}$ can be written as
follows
\begin{equation}\label{16}
\mathcal{P}^{\gamma}_{~\phi\hslash}=-\frac{1}{2}\mathcal{L}^{\gamma}_{~\phi\hslash}+
\frac{1}{4}(\mathcal{Q}^{\gamma}-\tilde{\mathcal{Q}}_{\gamma})g_{\phi\hslash}
-\frac{1}{4}\delta^{\gamma}~_{(\phi}\mathcal{Q}_{\hslash)}.
\end{equation}
This yields the relationship for $\mathcal{Q}$ as follows \cite{4-G}
\begin{equation}\label{17}
\mathcal{Q}=-\mathcal{Q}_{\gamma\phi\hslash}\mathcal{P}^{\gamma\phi\hslash}=
-\frac{1}{4}\bigg[-\mathcal{Q}^{\gamma\phi\hslash}
\mathcal{Q}_{\gamma\phi\hslash}+2\mathcal{Q}
^{\gamma\phi\hslash}\mathcal{Q}_{\hslash\gamma\phi}
-2\mathcal{Q}^{\varphi}\tilde{\mathcal{Q}_{\varphi}}
+\mathcal{Q}^{\varphi}\mathcal{Q}_{\varphi}\bigg].
\end{equation}
The field equations lead to the following formulations
\begin{eqnarray}\nonumber
\mathbb{T}_{\gamma\eta}&=&-\frac{1}{2}f(\mathcal{Q}, \mathbb{T})g_{\gamma\eta}
-\frac{2}{\sqrt -g}\nabla^{\phi}(f_\mathcal{Q}\sqrt-g\mathcal{P}_{\phi\gamma\eta})
-f_\mathcal{Q}(\mathcal{P}_{\phi\gamma\hslash}\mathcal{Q}^{~~\phi\hslash}_{\eta}
\\\label{19}
&-&2\mathcal{Q}^{\phi\hslash}_{\quad\gamma}\mathcal{P}_{\phi\hslash\eta})
+f_{\mathbb{T}}(\mathbb{T}_{\gamma\eta}+\theta_{\gamma\eta}),
\end{eqnarray}
where $f_\mathcal{Q} = \frac{\partial f}{\partial \mathcal{Q}}$ and $f_{\mathbb{T}} =
\frac{\partial f}{\partial \mathbb{T}}$. The variation $\delta\mathbb{T} =
\delta(\mathbb{T}_{\gamma\eta}g^{\gamma\eta}) = (\mathbb{T}_{\gamma\eta} +
\theta_{\gamma\eta})\delta g^{\gamma\eta}$ is used.

We will now examine the spherically symmetric metric for a general
spacetime. In spherical coordinates, the metric is expressed in the
following standard form
\begin{equation}\label{1a}
ds^{2}=-e^{\alpha(r)}dt^{2}+e^{\beta(r)}dr^{2}+r^{2}(d\theta^{2}+\sin^{2}\theta
d\phi^{2}).
\end{equation}
To describe the fluid distribution, we consider the anisotropic EMT as
\begin{equation}\label{B}
\mathbb{T}_{\gamma\eta} = (\varrho + p_t) u_\gamma u_\eta + p_t
g_{\gamma\eta} + (p_r - p_t) v_\gamma v_\eta,
\end{equation}
where $\varrho$ is the energy density, $p_r$ and $p_t$ represent the
radial and tangential pressures, respectively. In this framework,
the four-velocity components satisfy $u^\gamma u_\gamma = -1$, while
the unit radial four-vector satisfies $v^\gamma v_\gamma = 1$. In
$f(\mathcal{Q},\mathbb{T})$ gravity, the field equations are
expressed as \cite{11aa}-\cite{11gg}
\begin{align}\nonumber
\varrho &=-f_{\mathcal{Q}} \bigg(\mathcal{Q}(r)+\frac{1}{r^2}+\frac{
\big(\alpha '(r)+\beta '(r)\big)e^{-\beta (r)}}{r}\bigg)+\frac{2
f_{\mathcal{Q}\mathcal{Q}}  \mathcal{Q}'(r)e^{-\beta (r)}}{r}
\\\label{1c}
&-\frac{1}{3} f_{\mathbb{T}} (p_{r}+2 p_{t}+3 \varrho)+\frac{f}{2},
\\\label{1d}
p_{r}&=f_{\mathcal{Q}}
\big(\mathcal{Q}(r)+\frac{1}{r^2}\big)+\frac{2}{3} f_{\mathbb{T}}
(p_{t}-p_{r})-\frac{f}{2},
\\\nonumber
p_{t}&=f_{\mathcal{Q}} \bigg[\frac{\mathcal{Q}(r)}{2}-e^{-\beta (r)}
\big(\frac{\alpha ''(r)}{2}+\big(\frac{\alpha '(r)}{4}+\frac{1}{2
r}\big) \big(\alpha '(r)-\beta '(r)\big)\big)\bigg]
\\\label{1e}
&-f_{\mathcal{Q}\mathcal{Q}} e^{-\beta (r)} \mathcal{Q}'(r)
\bigg(\frac{\alpha '(r)}{2}+\frac{1}{r}\bigg)+\frac{1}{3}
f_{\mathbb{T}} (p_{r}-p_{t})-\frac{f}{2}.
\end{align}
Additionally, $\mathcal{Q}$ can be represented as follows \cite{11a}
\begin{equation}\label{1b}
\mathcal{Q}(r)=\frac{2e^{-\beta(r)}}{r}\bigg(\frac{1}{r}+\alpha '(r)\bigg).
\end{equation}
Here, the prime symbol signifies differentiation with respect to
$r$. We employ a specific model for $f(\mathcal{Q}, \mathbb{T})$,
expressed as \cite{10-b}
\begin{equation}\label{2a}
f(\mathcal{Q},\mathbb{T})=\mu  \mathcal{Q}+\nu  \mathbb{T},
\end{equation}
where, $\mu$ and $\nu$ are arbitrary non-zero constants. This
cosmological model is extensively reviewed in the current literature
\cite{10-c}. Substituting Eqs.\eqref{1b} and \eqref{2a} into
Eqs.\eqref{1c} through \eqref{1e}, we obtain
\begin{align}\nonumber
\varrho&=\frac{1}{36 (\nu +1) r^2}\{\mu  e^{-\beta (r)} \big(r
\big(2 (2 \nu -3) r \alpha ''(r)-(2 \nu -3) \alpha '(r) \big(r \beta
'(r)-4\big)
\\\label{2b}
&+(2 \nu -3) r \alpha '(r)^2-4 (\eta +6) \beta '(r)\big)-4 (\nu a
+6) \big(e^{\beta (r)}-1\big)\big)\},
\\\nonumber
p_{r}&=\frac{1}{36 (\nu +1) r^2}\{\mu  e^{-\beta (r)} \big(4 (\nu
+6) \big(e^{\beta (r)}-1\big)-r \big(2 (2 \nu -3) r \alpha
''(r)+\alpha '(r)
\\\label{2c}
&\times\big(8 (\nu +3)+(3-2 \nu ) r \beta '(r)\big)+(2 \nu -3) r
\alpha '(r)^2-4 (\nu -3) \beta '(r)\big)\big)\},
\\\nonumber
p_{t}&=\frac{1}{18 (\nu +1) r^2}\{\mu  e^{-\beta (r)} \big(r \big(-2
(\nu +3) r \alpha ''(r)+\alpha '(r) \big(-4 \nu +(\nu +3)r \beta
'(r)
\\\label{2d}
&-3\big)+(\nu +3) (-r) \alpha '(r)^2+(2 \nu +3) \beta '(r)\big)+2
(\nu -3) \big(e^{\beta (r)}-1\big)\big)\}.
\end{align}

We utilize the Finch-Skea solution because it provides a simple and
efficient model for obtaining viable outcomes for the interior
spacetime. This is due to its well-behaved nature and its ability to
satisfy all criteria for physical acceptability, as confirmed by
Delgaty and Lake \cite{x1}. Over the time, the Finch-Skea isotropic
stellar model has been extensively generalized by various
researchers to study a wide range of stellar objects. Additionally,
the Finch-Skea ansatz has been applied to interpret astrophysical
systems within different gravitational frameworks. In our current
model, the Finch-Skea ansatz and its solution play a central role.
Specifically, the form of $e^\beta$ has been directly adopted from
the Finch-Skea solution \cite{38}. However, the expression for
$e^\alpha$ includes modifications that were introduced in prior
studies \cite{x2} and are aligned with the framework of the
Finch-Skea metric used in this work. These modifications were
necessary to adapt the solution to the physical conditions and
requirements of the modified gravitational theory employed in this
study. Its regularity at the center and smooth matching between
interior and exterior regions make it ideal for analyzing compact
objects like gravastars. This solution is particularly useful in
studying the physical and geometric properties of gravastars within
the framework of modified gravity theories. These solutions are
formulated as \cite{x2}
\begin{equation}\label{3a}
e^{\alpha(r)}=\left(\xi
+\frac{1}{2}r\varphi\sqrt{r^2\chi}\right)^2,\quad e^{\beta(r)}=r^2
\chi +1.
\end{equation}
Here, $\xi$, $\varphi$ and $\chi$ are non-zero constants that can be
determined through matching conditions. By applying Eq.\eqref{3a}
into Eqs.\eqref{2b} through \eqref{2d}, we obtain the corresponding
equations
\begin{align}\nonumber
\varrho&=-\{\mu  \chi \big((\nu +6) r^5 \varphi  \chi ^2-5 (\nu -6)
r^3 \varphi  \chi +2 (\eta +6) \xi  \big(r^2 \chi \big)^{3/2}+6
(\eta +6)
\\\label{3b}
&\times \xi  \sqrt{r^2 \chi }+6 (3-2 \nu ) r \varphi \big)\}\{9 (\nu
+1) \big(r^2 \chi +1\big)^2 \big(r^3 \varphi \chi +2 \xi \sqrt{r^2
\chi }\big)\}^{-1},
\\\nonumber
p_{r}&=\{\mu  \chi  \big((\nu +6) r^5 \varphi  \chi ^2-(5 \nu +24)
r^3 \varphi  \chi +2 (\nu +6) \xi  \big(r^2 \chi \big)^{3/2}+6 \nu
\xi \sqrt{r^2 \chi }
\\\label{3c}
&-6 (2 \nu +3) r \varphi \big)\}\{9 (\nu +1) \big(r^2 \chi +1\big)^2
\big(r^3 \varphi \chi +2 \xi \sqrt{r^2 \chi }\big)\}^{-1},
\\\nonumber
p_{t}&=\{\mu  \chi  \big((\nu -3) r^5 \varphi  \chi ^2-(5 \nu +6)
r^3 \varphi  \chi +2 (\nu -3) \xi  \big(r^2 \chi \big)^{3/2}+6 \nu
\xi \sqrt{r^2 \chi }
\\\label{3d}
&-6 (2 \nu +3) r \varphi \big)\}\{9 (\nu +1) \big(r^2 \chi +1\big)^2
\big(r^3 \varphi \chi +2 \xi \sqrt{r^2 \chi }\big)\}^{-1}.
\end{align}

Anisotropic $(\Delta)$ pressure refers to the condition where the
pressure within a material or system is not uniform in all
directions. This directional dependence of pressure can
significantly influence the behavior and stability of physical
systems, such as stars, where anisotropic pressure can affect their
structural integrity and evolution \cite{10-f}. Figure \textbf{1}
shows that $p_r$ exceeds $p_t$ in the current model, leading to
negative anisotropy and an inward directed attractive force. Let us
consider $p_{r}$ and $p_{t}$ as $p$. Adding Eqs.\eqref{3c} and
\eqref{3d}, it follows that
\begin{eqnarray}\nonumber
p&=\{\mu  \chi  \big((2 \nu +3) r^5 \varphi  \chi ^2-10 (\nu +3) r^3
\varphi  \chi +2 (2 \nu +3) \xi  \big(r^2 \chi \big)^{3/2}+12 \nu
\xi
\\\label{4a}
&\times\sqrt{r^2 \chi }-(\nu +3) r \varphi \big)\}\{9 (\nu
+1)\big(r^2 \chi +1\big)^2 \big(r^3 \varphi \chi +2 \xi \sqrt{r^2
\chi }\big)\}^{-1}.
\end{eqnarray}
For mathematical analysis, treating $p_r$ and $p_t$ as a single $p$
makes the equations more manageable. This approach helps to simplify
the calculation of essential properties like stability and entropy
without compromising the gravastar core physical characteristics.
\begin{figure}\center
\epsfig{file=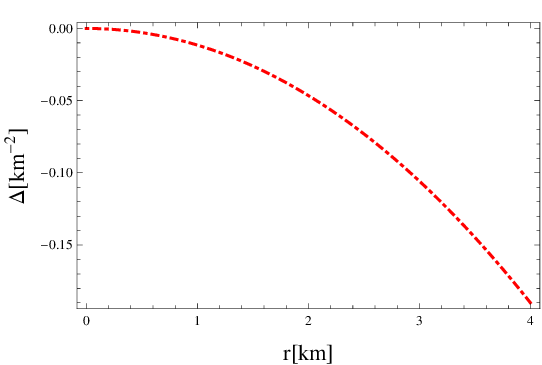,width=.5\linewidth}\caption{Plot of $\Delta$ against $r$.}
\end{figure}

\section{Gravastar Structure and Configuration}

In this section, we analyze the gravastar, focusing on three
distinct regions.
\begin{itemize}
\item
The core region, where the equation $p = -\varrho$ applies.
\item
The thin-shell region, where the relationship $p = \varrho$ is
valid.
\item
The outer region, where $p = 0$ is applicable.
\end{itemize}

\subsection{Inner Region}

The key parameter in the cosmic EoS is expressed as
$\omega=\frac{p}{\varrho}$, where $\omega$ is a variable applied
across the three distinct regions in the Mazur and Motola model
\cite{m}. In this scenario, we explore an intriguing gravitational
source in the inner region. Although dark matter and DE are commonly
regarded as separate entities, they might actually be different
manifestations of the same underlying phenomenon. Our aim is to
investigate the EoS to characterize the dark sector within the inner
region, as described by
\begin{equation}\label{5a}
p=-\varrho.
\end{equation}
By substituting Eq.\eqref{5a} into Eq.\eqref{3b} and then combining
the result with Eq.\eqref{4a}, we get
\begin{align}\nonumber
&\{\mu  \chi  \big((\nu -3) r^5 \varphi  \chi ^2-5 (\nu +12) r^3
\varphi  \chi +2 (\nu -3) \xi  \big(r^2 \chi \big)^{3/2}+6 (\nu -6)
\xi  \sqrt{r^2 \chi}
\\\label{6a}
&-6 (2 \nu +9) r \varphi \big)\}\{(\nu +1) \big(r^2 \chi +1\big)
\big(r^3 \varphi  \chi +2 \xi \sqrt{r^2 \chi }\big)\}^{-1}=0.
\end{align}
The active gravitational mass $\mathbf{M}(r)$ is determined as
\cite{10-g}
\begin{equation}\nonumber
\mathbf{M}(r)= 4 \pi  \int_0^r \zeta^2 \varrho (\zeta) d\zeta.
\end{equation}
Using Eq.\eqref{3b}, the result is given as follows
\begin{eqnarray}\nonumber
\mathbf{M}&=&4 \pi  \int \{r^2 \big(-\big(\mu  \chi \big((\nu +6)
r^5 \varphi  \chi ^2-5 (\nu -6) r^3 \varphi  \chi +2 (\nu +6)
\\\nonumber
&\times&\xi  \big(r^2 \chi \big)^{3/2}+6 (\eta +6) \xi \sqrt{r^2
\chi }+6 (3-2 \nu ) r \varphi \big)\big)\big)\}\{9 (\nu +1) \big(r^2
\chi +1\big)^2
\\\nonumber
&\times& \big(r^3 \varphi  \chi +2 \xi \sqrt{r^2 \chi }\big)\}^{-1}
\, dr.
\end{eqnarray}
In Figure \textbf{2}, the numerical solution to this equation is
shown, confirming the model dependability. The graph reveals a
continuous increase in mass as radius grows, which aligns with the
physical principle that mass should always rise as a function.
\begin{figure}\center
\epsfig{file=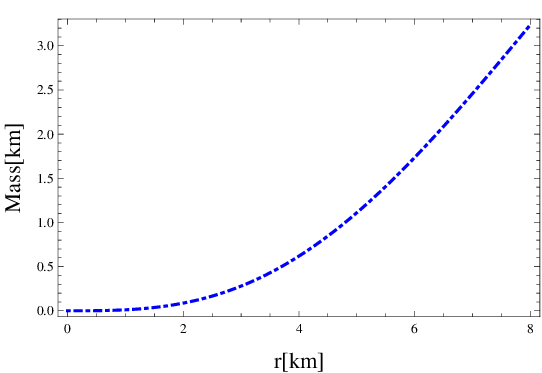,width=.48\linewidth} \caption{Plot of $\mathbf{M}$ with respect
to $r$.}
\end{figure}

\subsection{Shell}

Here, we analyze a stiff perfect fluid governed by the EoS defined
as
\begin{equation}\label{aaa}
p =\varrho.
\end{equation}
This fluid is confined within a thin-shell. The EoS used here is a
specific form of a barotropic EoS, defined by $\omega = 1$, which
leads to the relation $p = \omega \varrho$. In barotropic fluids,
pressure depends solely on density, expressed as $p = p(\varrho)$,
indicating a direct interdependence between pressure and density.
Although these types of fluids are often considered improbable, they
provide an essential educational resource by demonstrating various
techniques for investigating different systems and addressing
physically interesting challenges. In this context,
Zel$^{\prime}$dovich \cite{1b} was the first to introduce the idea
of this fluid, characterizing it as a stiff fluid. Staelens et al.
\cite{1c} conducted an in-depth analysis of the gravitational
collapse of an over-dense barotropic fluid governed by a linear EoS
within a cosmological framework. This paradigm has been extensively
adopted by the astrophysical and cosmological research communities
\cite{1d}. Solving the field equations in non-vacuum regions, such
as boundary layers, inherently presents significant computational
challenges. Nevertheless, an analytical solution was successfully
derived under the thin-shell approximation, specifically under the
condition that $e^{-\beta(r)}$ remains positive and markedly smaller
than unity. Building upon Israel conjecture \cite{2a}, we posit that
the intermediary region separating the two spacetime manifolds must
constitute a thin-shell. Furthermore, as the radial coordinate $r$
approaches to zero, any parameter dependent on $r$ can generally be
considered negligible (i.e., substantially less than one). This
approximation facilitates the simplification of our field equations
from Eqs.\eqref{1c}-\eqref{1e} to the following
\begin{align}\label{7a}
\varrho+ \frac{\nu}{2} (5p + \varrho) &=\mu \left( \frac{e^{-\beta(r)} \beta'(r)}{r}
+ \frac{1}{r^2} \right),
\\\label{7b}
p+\frac{\nu}{2} (\varrho - 3p)&=\mu \left( -\frac{1}{r^2} \right),
\\\label{7c}
p+ \frac{\nu}{2} (\varrho - 3p)&=-\mu \left( \beta'(r) \alpha'(r)
\frac{e^{-\beta(r)}}{4} + \frac{e^{-\beta(r)} \beta'(r)}{2r} \right).
\end{align}
Substituting Eqs.\eqref{1b}, \eqref{2a}, \eqref{3a} and \eqref{aaa}
into Eqs.\eqref{7a} through \eqref{7c}, we first combine
Eqs.\eqref{7a} and \eqref{7b}, and then add Eqs.\eqref{7b} and
\eqref{7c}. Solving the resulting pair of equations leads us to the
final outcome
\begin{equation}\label{A}
e^{-\beta (r)}=-c_{1}-\frac{(\nu +2) \left(r^2 \chi +1\right)
\left(2 \xi +r \varphi  \sqrt{r^2 \chi }\right)}{r^2 \chi  \left(6
(\nu -2) \xi +(\nu -10) r \varphi  \sqrt{r^2 \chi }\right)},
\end{equation}
where $c_{1}$ is the constant of integration. Figure \textbf{3}
depicts the variations in $p$ and $\varrho$. It is evident that
$\varrho$ within the shell steadily rises as it approaches the outer
boundary. Composed of a stiff fluid, the shell demonstrates that
both $p$ and $\varrho$ increase uniformly towards the outer surface.
This indicates that the concentration of stiff matter is greater at
the outer edge than in the shell inner region.
\begin{figure}\center
\epsfig{file=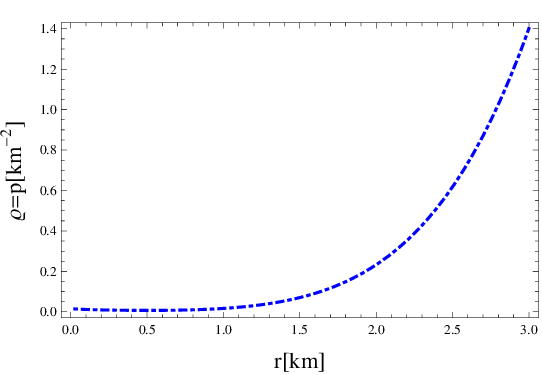,width=.48\linewidth} \caption{Plot of $\varrho=p$ with respect to
$r$.}
\end{figure}

\subsection{Exterior Region}

The gravastar exterior conforms to the EoS $p = 0$, implying that
the outer layer is completely surrounded by a vacuum. Among the most
intriguing solutions to the Einstein field equations in GR is the
Schwarzschild solution, which describes a spherically symmetric,
static vacuum. The line element of the Schwarzschild metric can be
expressed as follows
\begin{equation}\label{8a}
ds_+^{2} = -\left( 1 - \frac{2M}{r} \right) dt^2 + \left( 1 -
\frac{2M}{r} \right)^{-1} dr^2 + r^2(d\theta^2 + \sin^2\theta
d\phi^2).
\end{equation}
In this context, the symbol $+$ denotes the exterior solution. A
black hole with an isolated mass $M$, represented by the static
spherically symmetric line element \eqref{8a}, is called a
Schwarzschild black hole.

\section{Boundary and Junction Conditions}

In this section, we synchronize the internal spacetime with the
external spacetime to determine the constants $\chi$, $\xi$ and
$\varphi$. By ensuring the continuity of the metric coefficients
$g_{tt}$, $g_{rr}$ and $\frac{\partial g_{tt}}{\partial r}$, we
establish a seamless connection between the interior and exterior
regions at the boundary surface $r = R$. This matching procedure is
crucial for maintaining the smoothness and consistency of the
spacetime geometry at the interface between the two domains. We
proceed by calculating the values of $\chi$, $\xi$ and $\varphi$
specifically for the Schwarzschild black hole configuration. In this
process, we juxtapose the temporal and radial metric coefficients
from Eqs.\eqref{1a} and \eqref{8a}. Applying Eq.\eqref{3a}, we
ascertain that
\begin{eqnarray}\label{10a}
1-\frac{2 M}{R}&=&\bigg(\xi +\frac{1}{2} R \varphi \sqrt{R^2 \chi
}\bigg)^2,
\\\label{10b}
\bigg(1-\frac{2 M}{R}\bigg)^{-1}&=&R^2 \chi +1.
\end{eqnarray}
Differentiating Eq.\eqref{10a} with respect to $R$, we have
\begin{equation}\label{10c}
\frac{2 M}{R^2}= \varphi \bigg(R^3 \varphi \chi +2 \xi \sqrt{R^2
\chi }\bigg).
\end{equation}
The constants $\chi$, $\xi$ and $\varphi$ are expressed in terms of $M$ and $R$
through Eqs.\eqref{10a}-\eqref{10c}. Applying these equations, we determine the
following values
\begin{align}\label{1aa} \chi &= \frac{2M}{R^2(R
- 2M)},
\\\label{2aa}
\xi& = \pm \frac{\sqrt{\frac{M}{R - 2M}}(2R - 5M)}{2\sqrt{M}\sqrt{R}},
\\\label{3aa}
\varphi&=\pm \frac{\sqrt{M}}{\sqrt{2} R^{3/2}}.
\end{align}
During the derivation, we accounted for both the positive and
negative branches of $\xi$ and $\varphi$ arising from the square
root. To ensure the model physical viability and consistency, we
select the positive branch, aligning with established practices in
the literature, where only positive values are considered to
maintain plausibility.
\begin{eqnarray}\label{10d}
\chi&=&\frac{2 M}{R^2 (R-2 M)},
\\\label{10e}
\xi&=&\frac{\sqrt{\frac{M}{R-2 M}} (2 R-5 M)}{2 \sqrt{M} \sqrt{R}},
\\\label{10f} \varphi&=&\frac{\sqrt{M}}{\sqrt{2} R^{3/2}}.
\end{eqnarray}

The exploration of junction conditions, which govern the seamless
integration between two distinct regions such as a star interior and
exterior, was initiated by Sen \cite{2b}. These conditions are
essential for ensuring smooth transitions across different spacetime
domains. The Darmois junction conditions \cite{2d} facilitate this
seamless connection by enforcing two main criteria: the interface
surface must preserve the same geometry and dimensions on both sides
(metric continuity), and the curvature of this surface must remain
consistent across both regions (extrinsic curvature continuity). In
contrast, the Israel junction conditions \cite{2a} are applicable
when a thin layer, like a membrane, exists at the boundary, carrying
its own energy and pressure. In such cases, the Israel conditions
permit a sudden change in surface curvature across the layer,
reflecting the unique properties of the thin boundary. In this work,
we employ the Israel junction conditions \cite{3a} due to their
effectiveness in scenarios involving a boundary between two distinct
spacetime regions separated by a thin-shell of matter or energy.
This methodology is especially suitable for examining surface layers
that carry mass. While the metric coefficients remain continuous
across the boundary, their derivatives may show discontinuities at
the hypersurface, which is consistent with the requirements for
analyzing such thin-shells.

The metric induced on the boundary is represented by the following
line element
\begin{equation}\label{2ab}
ds^2 = -d\tau^2 + R(\tau)^2 d\theta^2 + R(\tau)^2 \sin^2\theta \, d\phi^2.
\end{equation}
Here, $\tau$ represents the proper time. To maintain the stability of the thin-shell,
it is necessary to apply the Lanczos equations to determine the surface tension and
pressure \cite{33aa}
\begin{equation}\label{50}
S^{i}_{j}= -\frac{1}{8\pi}\{[A^{i}_{j}]-\delta^{i}_{j}A\},
\end{equation}
where $[A^{i}_{j}] = A^{+i}_{j}-A^{-i}_{j}$, $A=$~tr$[A_{ij}]=[A^{i}_{j}]$ and the
indices $i$ and $j$ correspond to the coordinates on the hypersurface $(i, j = 0, 2,
3)$. The extrinsic curvature is defined as follows
\begin{equation}\label{48}
A_{ij}^{\pm}=-n_{v}^{\pm}\bigg(\frac{d^{2}x^{v}_{\pm}} {d\phi^{i}d\phi^{j}}+
\Gamma_{lm}^{v}\frac{d
x^{l}_{\pm}}{d\phi^{i}}\times\frac{dx^{m}_{\pm}}{d\phi^{j}}\bigg),\quad l,m=0,1,2,3.
\end{equation}
Here, $\phi^{i}$ represents the intrinsic coordinates and the
components of $A^{i\pm}_{j}$ are defined as follows
\begin{equation}\label{4aab}
A^{\tau\pm}_{\tau} = \frac{\Pi^\prime + 2\ddot{R}}{\sqrt{\Pi_{\pm}(R) + \dot{R}^2}},
\quad A^{\theta\pm}_{\theta} = \frac{\sqrt{\Pi_{\pm}(R) + \dot{R}^2}}{R}, \quad
A^{\phi\pm}_{\phi} = \sin^2\theta \, A^{\theta\pm}_{\theta}.
\end{equation}
The unit normals to the hypersurface $n_{v}^{\pm}$ are expressed as
follows
\begin{equation}\label{49}
n_{v}^{\pm}=\bigg(\frac{\dot{R}}{\Pi_{\pm}(R)},\sqrt{\Pi_{\pm}(R)
+\dot{R}^{2}},0,0\bigg),
\end{equation}
where $\dot{R}$ represents derivative of $R$ with respect to $\tau$.

The Lanczos equations are employed to determine the surface energy density $\rho$ and
surface pressure $P$ of thin-shell gravastars, which are defined as follows
\begin{eqnarray}\label{4aac}
\rho&=& -\frac{[A^\theta_\theta]}{4\pi} = -\frac{1}{4\pi R}\left(\sqrt{\dot{R}^2 +
\Pi_{+}(R)} - \sqrt{\dot{R}^2 + \Pi_{-}(R)}\right),
\\\nonumber
P&=& \frac{[A^\theta_\theta] + [A^\tau_\tau]}{8\pi}= \frac{1}{8\pi
R}\bigg[\frac{2\dot{R}^2 + 2R\ddot{R} + 2\Pi_{+}(R) +
R\Pi^{\prime}_{+}(R)}{\sqrt{\dot{R}^2 + \Pi_{+}(R)}}
\\\label{4aad}&-&\frac{2\dot{R}^2 + 2R\ddot{R} + 2\Pi_{-}(R) +
R\Pi^{\prime}_{-}(R)}{\sqrt{\dot{R}^2 + \Pi_{-}(R)}}\bigg].
\end{eqnarray}
When the system reaches equilibrium, with the shell remaining fixed at the throat
radius $R = R_{0}$ (i.e., $\dot{R_{0}} = 0$ and $\ddot{R_{0}}=0$), the equations
reduce to the following form
\begin{eqnarray}\label{13f}
\rho_{0}&=&-\frac{1}{4\pi R_{0}}[\sqrt{\Pi_{+}(R_{0})}-\sqrt{\Pi_{-}(R_{0})}],
\\\label{13g}
P_{0}&=&-\rho_{0}
+\frac{1}{8\pi}\bigg[\frac{\Pi_{+}^{\prime}(R_{0})}{\sqrt{\Pi_{+}(R_{0})}}-
\frac{\Pi_{-}^{\prime}(R_{0})}{\sqrt{\Pi_{-}(R_{0})}}\bigg].
\end{eqnarray}
The exterior function is expressed as
\begin{equation}\label{52a}
\Pi_{+}(R_{0})=1-\frac{2 M}{R_{0}}.
\end{equation}
The interior function is characterized as
\begin{eqnarray}\nonumber
\Pi_{-}(R_{0})&=&-c_{1}-\{(\nu +2) R_{0}^2 (R_{0}-2 M) \bigg(\frac{2 R_{0}^2
M}{R_{0}^2 (R_{0}-2 M)}+1\bigg) \big(\sqrt{2} R_{0} \psi
\\\nonumber
&\times&\{\{R_{0}^2 M\}\{R_{0}^2 (R_{0}-2
M)\}^{-1}\}^{\frac{1}{2}}+\frac{\sqrt{\frac{M}{R_{0}-2 M}} (2 R_{0}-5 M)}{\sqrt{M}
\sqrt{R_{0}}}\big)\}\{2 R_{0}^2 M
\\\nonumber
&-&\big(\sqrt{} R_{0}(\nu) \psi \sqrt{\frac{R_{0}^2 M}{R_{0}^2 (R_{0}-2 M)}}+\{ (\nu
-2) \sqrt{\frac{M}{R_{0}-2 M}} (R_{0}-5 M)\}
\\\label{52b}
&\times&\{\sqrt{M} \sqrt{R_{0}}\}\big)\}^{-1}.
\end{eqnarray}

The dynamic behavior of a thin-shell can be analyzed through the
equation of motion and the conservation principle. These equations
are fundamental for determining the stable regions within a
geometric configuration. The equation of motion can be derived by
rearranging Eq.\eqref{4aac} as follows
\begin{equation}\label{5abc}
\dot{R}^2 + V(R) = 0.
\end{equation}
The effective potential, denoted by $V(R)$, can be expressed as follows \cite{44aa}
\begin{eqnarray}\label{13j}
V(R)&=\frac{1}{2} \big( \Pi_{-}(R) + \Pi_{+}(R) \big) - \frac{\big(\Pi_{-}(R) -
\Pi_{+}(R) \big)^2}{64\pi^2 R^2 \rho^2} - 4\pi^2 R^2 \rho^2.
\end{eqnarray}
Next, we examine that $\rho$ and $P$ satisfy the conservation equation
\begin{equation}\label{59}
\frac{d}{d\tau}(\rho 4\pi R^2)+P\frac{d 4\pi R^2}{d\tau}=0.
\end{equation}
By utilizing this equation, we are able to determine
\begin{equation}\label{60}
\rho^{\prime}=-(\rho+P)\frac{2}{R}.
\end{equation}

Now, we analyze the behavior of $\rho_{0}$ and $P_{0}$ at the junction surface
$R=R_{0}$. Substituting Eqs.\eqref{52a} and \eqref{52b} into Eqs.\eqref{13f} and
\eqref{13g}, we obtain
\begin{eqnarray}\nonumber
\rho_{0}&=&\frac{1}{8 \pi  R_{0}}\bigg[2 \sqrt{1-\frac{2 M}{R_{0}}}-\sqrt{2}(\{2
\sqrt{2} R_{0}^{3/2}c_{1} (\nu -10) M^{3/2} \psi +\sqrt{2} R_{0}^{5/2} (\nu +2)
\\\nonumber
&\times& \sqrt{M} \psi +2 R_{0}^2 (\nu +2)+R_{0} M (12 c_{1} (\nu -2)-5 (\nu
+2))-30c_{1} (\nu -2) M^2\}
\\\label{14a}
&\times&\{M \big(-\sqrt{2} R_{0}^{3/2} (\nu -10) \sqrt{M} \psi -6 R_{0} (\nu -2)+
(\nu -2) M\big)\}^{-1})^{\frac{-1}{2}}\bigg],
\\\nonumber
P_{0}&=&\frac{1}{8 \pi }\bigg[\frac{M}{R_{0}^2 \sqrt{1-\frac{2
M}{R_{0}}}}+\frac{1}{R_{0}}\bigg\{2 \sqrt{1-\frac{2 M}{R_{0}}}-\sqrt{2} \{\{2
\sqrt{2} R_{0}^{3/2}c_{1} (\nu -10) M^{3/2} \psi
\\\nonumber
&+&\sqrt{2} R_{0}^{5/2} (\nu +2) \sqrt{M} \psi +2 R_{0}^2 (\nu +2)+R_{0} M (12 c_{1}
(\nu -2)-5 (\nu +2))
\\\nonumber
&-&30 c_{1} (\nu -2) M^2\}\{M \big(-\sqrt{2} R_{0}^{3/2} (\nu -10) \sqrt{M} \psi
-6R_{0} (\nu -2)+15 (\nu -2)
\\\nonumber
&\times& M\big)\}^{-1}\}^{\frac{1}{2}}\bigg\}+\{(\nu +2) \big(5 \sqrt{2} R_{0}^{3/2}
(10-7 \nu ) M^{3/2} \psi +2 \sqrt{2} R_{0}^{5/2} (5 \nu -14)
\\\nonumber
&\times& \sqrt{M} \psi +2 R_{0}^3 (\nu -10) M \psi ^2+12 R_{0}^2 (\nu -2)-60 R_{0}
(\nu -2) M+75 (\nu -2)
\\\nonumber
&\times& M^2\big)\}\{2 \sqrt{2} M \big(\sqrt{2} R_{0}^{3/2} (\nu -10) \sqrt{M} \psi
+6 R_{0} (\nu -2)-15 (\nu -2) M\big)^2
\\\nonumber
&\times& \{\{2 \sqrt{2} R_{0}^{3/2} c_{1} (\nu -10) M^{3/2} \psi +\sqrt{2}
R_{0}^{5/2} (\nu +2) \sqrt{M} \psi +2 R_{0}^2 (\nu +2)
\\\nonumber
&+&R_{0} M (12 c_{1} (\nu -2)-5 (\nu +2))-30 c_{1} (\nu -2) M^2\}\{M \big(-\sqrt{2}
R_{0}^{3/2} (\nu -10)
\\\label{14b}
&\times& \sqrt{M} \psi -6 R_{0} (\nu -2)+15 (\nu -2)
M\big)\}^{-1}\}^{\frac{1}{2}}\}^{-1}\bigg].
\end{eqnarray}
Figure \textbf{4} illustrates the variation of surface energy density and pressure
with the thickness of thin-shell. The graph demonstrates that the energy density is
positive but gradually declines from the inner boundary to the outer boundary.
Meanwhile, the pressure remains negative and also decreases as it nears the outer
boundary.
\begin{figure}\center
\epsfig{file=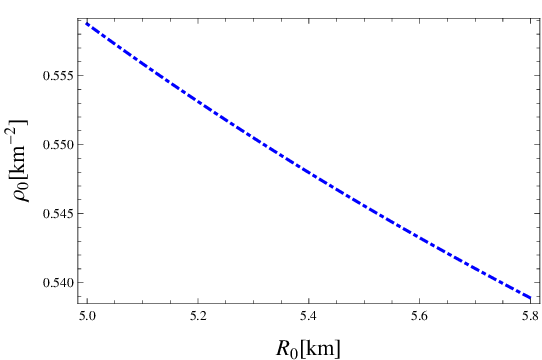,width=.48\linewidth}
\epsfig{file=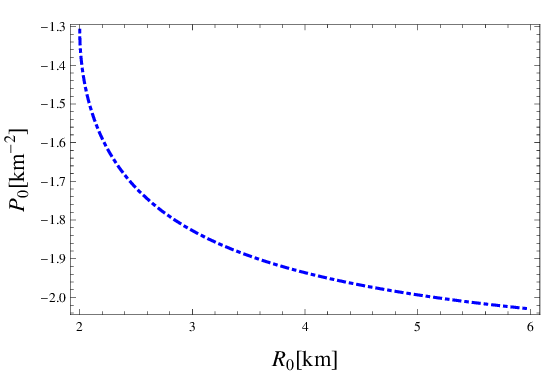,width=.46\linewidth}\caption{Plots of $\rho_{0}$ and $P_{0}$ with
respect to $R_{0}$.}
\end{figure}

\section{Physical Features of Gravastar}

In this section, we will examine the physical properties of a
gravastar, including its energy, proper length, energy conditions,
entropy and the EoS parameter, at the shell equilibrium.
Additionally, we will investigate the stability of the gravastar
thin shell, with a focus on the effective potential, redshift,
causality condition and adiabatic index.

\subsection{Energy}

The center of a gravastar indicates the existence of DE due to the negative pressure
that produces a repulsive force. The energy of the shell can be determined using the
following formula \cite{5aaa}
\begin{equation}
E = \int_{R_{0}}^{R_{0}+\epsilon} 4\pi R^2\rho_{0} dR.
\end{equation}
Using Eqs.\eqref{14a}, we deduce that
\begin{eqnarray}\nonumber
E&=&-\bigg[4 \pi  \zeta  \chi  \big((R_{0}+\epsilon )^3-R_{0}^3\big) \big((\eta +6)
R_{0}^5 \varphi \chi ^2-5 (\eta -6) R_{0}^3 \varphi  \chi +2 (\eta +6)
\\\nonumber
&\times&\xi  \big(R_{0}^2 \chi \big)^{3/2}+6 (\eta +6) \xi  \sqrt{R_{0}^2 \chi }+6
(3-2 \eta ) R_{0} \varphi \big)\bigg]\bigg[27 (\eta +1) \big(R_{0}^2 \chi +1\big)^2
\\\label{17}
&\times&\big(R_{0}^3 \varphi \chi+2 \xi \sqrt{R_{0}^2 \chi }\big)\bigg]^{-1}.
\end{eqnarray}
Figure \textbf{5} shows that the shell energy increases as its
thickness grows.
\begin{figure}\center
\epsfig{file=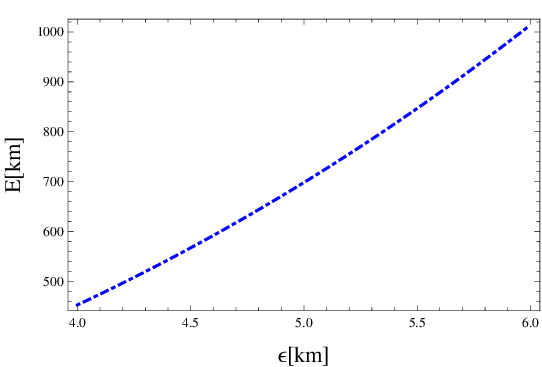,width=.48\linewidth} \caption{Plot of energy with respect to
thickness.}
\end{figure}

\subsection{Proper Length}

According to the models proposed by Mazur and Mottola \cite{m}, the shell is
positioned at the junction where two spacetimes intersect. The length of the shell
spans from the outer boundary $r_2 = R_{0} + \epsilon$ ($\epsilon$ denotes the
thickness of the shell and a small distance), which marks the transition from the
outer spacetime to the intermediate thin-shell, to the inner boundary $r_1 = R_{0}$,
which separates the inner region from the intermediate thin-shell. The thickness of
the shell calculated between these two phase boundaries is given by \cite{5aaa}
\begin{equation}
L= \int_{R_{0}}^{R_{0}+\epsilon} \sqrt{e^{\beta(R)}}dR.
\end{equation}
This can be converted into the interior function as
\begin{eqnarray}\nonumber
L&=&\int_{R_{0}}^{R_{0}+\varepsilon} \bigg[-c_{1}-\bigg[(\nu +2) R^2 (R-2 M)
\bigg(\frac{2 R^2 M}{R^2 (R-2 M)}+1\bigg) \big(\sqrt{2} R \psi
\\\nonumber
&\times&\{\{R^2 M\}\{R^2 (R-2 M)\}^{-1}\}^{\frac{1}{2}}+\frac{\sqrt{\frac{M}{R-2 M}}
(2 R-5 M)}{\sqrt{M} \sqrt{R}}\big)\bigg]\bigg[2 R^2 M \\\label{65} &-&\big( R(\nu)
\psi \sqrt{\frac{R^2 M}{R^2 (R-2 M)}}+\frac{ (\nu -2) \sqrt{\frac{M}{R-2 M}} (R-5
M)}{\sqrt{M} \sqrt{R}}\big)\bigg]^{-1}\bigg]d R.
\end{eqnarray}
Using Eq.\eqref{52b}, we obtain
\begin{equation}
= \int_{R_{0}}^{R_{0}+\epsilon} \frac{1}{\Pi_-(R)}dR.
\end{equation}
Calculating the integral in Eq.\eqref{65} proves to be quite challenging. To simplify
the process, we set $\frac{d\Pi_-(R)}{dR} = \frac{1}{\Pi_-(R)}$ to solve the
integral. This yields
\begin{equation}\label{16aa}
L = \Pi_-(R_{0} + \epsilon) - \Pi_-(R_{0}).
\end{equation}
Expanding $\Pi_-(R_{0} + \epsilon)$ in a Taylor series around $R_{0}$ and considering
only the linear terms in $\epsilon$. Since $\epsilon$ is very small, higher-order
terms in the expansion can be neglected. Figure \textbf{6} illustrates the
relationship between proper length and shell thickness, showing that as the thickness
increases, the proper length of the shell grows proportionally.
\begin{figure}\center
\epsfig{file=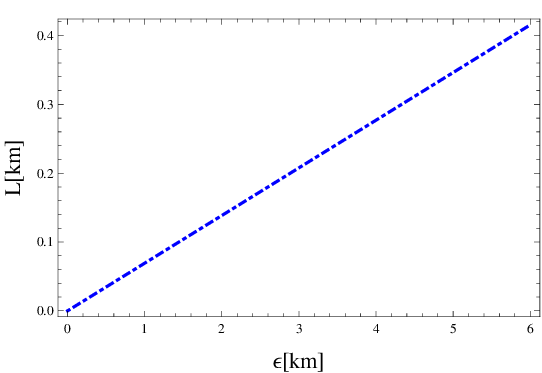,width=.48\linewidth} \caption{Plot of proper length with respect
to thickness.}
\end{figure}

\subsection{Energy Conditions}

To evaluate the physical viability of geometric structures in
modified theories of gravity, it is crucial to follow specific
criteria known as energy conditions. These conditions act as
essential guidelines to determine whether a particular arrangement
of matter and energy can exist within the framework of these
extended theories. By applying these criteria, we ensure that the
resulting structures are consistent with our understanding of
gravitational physics. The commonly recognized energy criteria are
\begin{itemize}
\item Null energy conditions (NEC) states that
$0 \leq \rho_{0} + P_{0}$.
\item Weak energy conditions is given as
$0 \leq \rho_{0} + P_{0},~ 0 \leq \rho_{0}$.
\item Strong energy conditions is formulated as
$0 \leq \rho_{0} + P_{0},~ 0 \leq \rho_{0} + 2P_{0}$.
\item Dominant energy conditions is given as
$0 \leq \rho_{0} \pm P_{0}$.
\end{itemize}
For the model to be considered physically viable, it must comply with these energy
conditions. Our primary focus is whether the NEC is met, as this criterion signifies
the existence of either ordinary or exotic matter within the thin-shell. It is
crucial to emphasize that any violation of the NEC could result in the infringement
of the other energy conditions. Figure \textbf{7} demonstrates that the NEC is
satisfied for the given model parameter, confirming the physical viability of the
model.
\begin{figure}\center
\epsfig{file=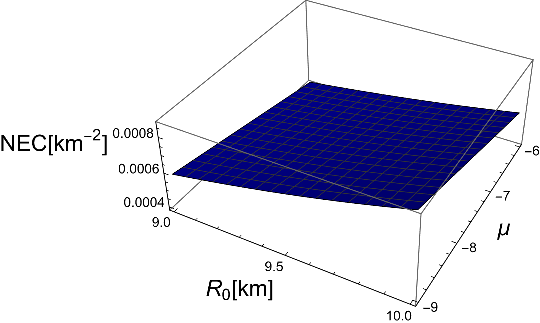,width=.48\linewidth} \caption{Plot of the NEC versus the model
parameter $\mu$.}
\end{figure}

\subsection{Entropy}

According to Mazur and Mottola \cite{m}, the entropy $S$ for the
intermediate thin-shell can be calculated using the following
formula
\begin{equation}\label{16bb}
S = \int_{R_{0}}^{R_{0}+\epsilon} 4\pi R^2 s(R) \sqrt{e^{\beta(R)}}dR.
\end{equation}
The entropy density $s(R)$ is defined as
\begin{equation}
s(R) = \frac{k^2 K_B^2 \mathrm{T}(R)}{4\pi \hbar^2 G} = \frac{k K_B}{\hbar}
\sqrt{\frac{P_{0}}{2\pi G}},
\end{equation}
where $k$ is non-zero arbitrary constant, $\mathrm{T}(R)$ represents
the temperature, $\hbar$ denotes Planck constant and $K_B$ is the
Boltzmann constant. In this analysis, we use geometrical units $G =
1$. Substituting this into Eq.\eqref{16bb}, the expression for $S$
becomes
\begin{eqnarray}\nonumber
S &=& \int_{R_{0}}^{R_{0}+\varepsilon} 4 \pi R^{2} \frac{k
K_{B}}{\hbar}(\frac{P_{0}}{2 \pi})^{\frac{1}{2}}  \sqrt{e^{\beta(R)}} dR=
\int_{R_{0}}^{R_{0}+\varepsilon} 2 \sqrt{2 \pi} R^{2} \frac{k K_{B}}{\hbar}
\sqrt{P_{0} e^{\beta(R)}} d R.
\end{eqnarray}
This can be simplified to
\begin{equation}\label{16b}
S = 2\sqrt{2\pi}R^2 k\frac{ K_B}{\hbar} N,
\end{equation}
where
\begin{equation}\label{16c}
N = \int_{R_{0}}^{R_{0}+\epsilon}\sqrt{P_{0}
e^{\beta(R)}}dR=\int_{R_{0}}^{R_{0}+\epsilon} D(R)dR.
\end{equation}
Evaluating the integral in Eq.\eqref{16c} is quite complex at this stage. To
simplify, let us consider $F(R)$ as the antiderivative of $D(R)$. This allows us to
compute the integral using the fundamental theorem of calculus, which modifies
Eq.\eqref{16c} to
\begin{equation}\label{16d}
N = [F(R)]_{R_{0}}^{R_{0}+\epsilon} = F(R_{0} + \epsilon) - F(R_{0}).
\end{equation}
Expanding $F(R_{0} + \epsilon)$ in a Taylor series around $R_{0}$
and keeping only the first-order term in $\epsilon$, we derive from
Eq.\eqref{16b}. However, due to the complexity of the equations, we
will present only their graphical representations. Figure \textbf{8}
demonstrates that as the shell thickness increases, the disorder
within the gravastar rises, with entropy increasing in proportion to
the shell thickness and peak at the outer surface. A shell with no
thickness results in zero entropy.
\begin{figure}\center
\epsfig{file=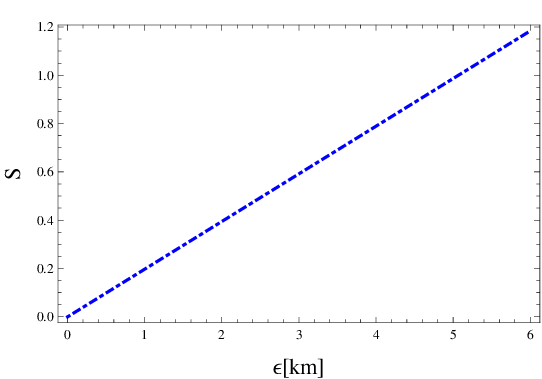,width=.48\linewidth} \caption{Plot of entropy with respect to
thickness.}
\end{figure}

\subsection{The EoS Parameter}

The EoS parameter, denoted as $\omega$, is vital in cosmology and
thermodynamics, relating $P_{0}$ to $\rho_{0}$ of a cosmic fluid
through the equation $P_{0} = \omega \rho_{0}$ \cite{11}. It
provides insight into the behavior of different energy forms in the
universe: $\omega = 0$ corresponds to non-relativistic matter,
$\omega = \frac{1}{3}$ describes radiation, and DE typically has
$\omega < -\frac{1}{3}$. Observational data from cosmic microwave
background radiation, galaxy clusters and supernovae suggest that DE
has a value close to $\omega \approx -1$, while $\omega < -1$
corresponds to phantom energy and $\omega \geq -1$ defines the
non-phantom regime. Current research explores potential variations
in the EoS parameter over time and the implications of different
forms of DE, making it essential for understanding the universe fate
and the evolution of its energy components. Figure \textbf{9}
illustrates that $\omega$ falls below -1, entering the phantom
region and highlighting the importance of the gravastar structure.
\begin{figure}\center
\epsfig{file=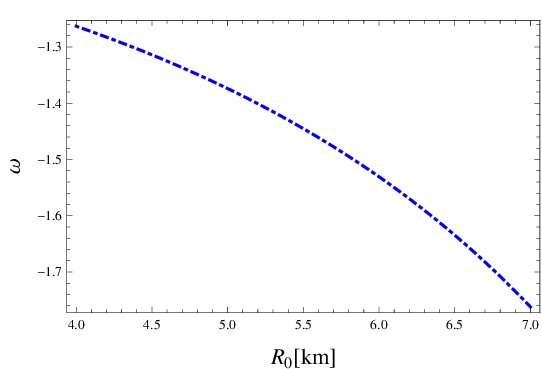,width=.48\linewidth} \caption{Plot of EoS with respect to
$R_{0}$.}
\end{figure}

\subsection{Stability Analysis}

Here, we will investigate the stability of the gravastars at the shell equilibrium,
with a focus on the effective potential. A positive $V^{\prime\prime}(R_{0})$
indicates a stable structure, while a negative value points to instability. When
$V^{\prime\prime}(R_{0})$ equals to zero, the stability cannot be predicted
\cite{44aa}. We present only the graphical behavior of $V^{\prime\prime}(R_{0})$ due
to the complexity of the expressions. Figure \textbf{10} confirms that
$V^{\prime\prime}(R_{0}) > 0$ ensuring the stable configuration of the thin-shell
gravastar.
\begin{figure}\center
\epsfig{file=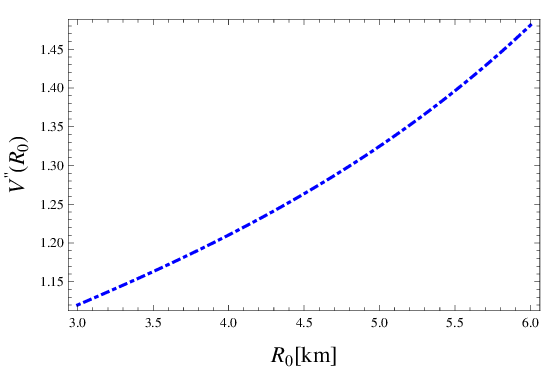,width=.47\linewidth} \caption{Plot of $V^{\prime\prime}(R_{0})$
against $R_{0}$.}
\end{figure}

Gravitational redshift, denoted by $Z(R)$, refers to the phenomenon
where light emitted from a gravitational field experiences a shift
to longer wavelengths as it escapes a massive object gravitational
influence. This occurs because gravity impacts the frequency of
photons as light moves upwards from a gravitational well, its energy
decreases, causing an increase in wavelength. The mathematical
expression for gravitational redshift is given by
\begin{equation*}
Z(R) = \frac{1}{\sqrt{-g_{tt}}}-1.
\end{equation*}
Additionally, the surface redshift must remain below 2 for a stable,
isotropic distribution of perfect fluid \cite{62a}. This requirement
is critical for the stability of such systems, as it prevents
excessive gravitational forces that could lead to instability or
collapse. A surface redshift exceeding this limit suggests a
gravitational field strong enough to compromise the fluid integrity,
potentially leading to singularities or event horizons. Thus,
ensuring the surface redshift remains within this boundary is
essential for the stability and physical plausibility of perfect
fluid distributions in GR. In Figure \textbf{11}, the redshift
parameter stays within the expected range throughout the gravastar
shell, with $Z(R)$ remaining below 2. The gravitational redshift is
positive and finite, steadily increasing towards the surface. This
confirms the gravastar model physical stability and compatibility
with the $f(\mathcal{Q}, \mathbb{T})$ framework.
\begin{figure}\center
\epsfig{file=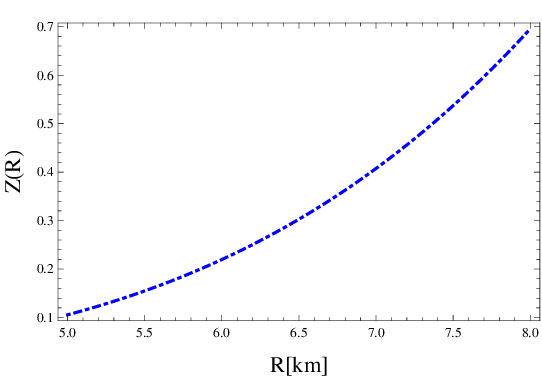,width=.47\linewidth} \caption{Plot of $Z(R)$ against $R$.}
\end{figure}

We also analyze the stability by examining the causality condition and the adiabatic
index. Having computed $\varrho$ and $p$ for the thin-shell from Eq.\eqref{7a} and
Eq.\eqref{7b}, we can express the speed of sound as
\begin{equation}\label{18}
\upsilon_{s}^{2} =\frac{p^{\prime}}{\varrho^{\prime}}.
\end{equation}
To determine the speed of sound specifically within the thin-shell, we evaluate this
expression at $R$. According to Poisson and Visser \cite{4a}, the causality condition
dictates that the speed of sound should not exceed the speed of light (where $c = 1$
in natural units). Therefore, the speed of sound $\upsilon_{s}^{2}$ is anticipated to
lie within the range $0 \leq \upsilon_{s}^{2} \leq 1$. Lobo \cite{4b} applied this
criterion to set a limit on wormhole stability in the presence of a cosmological
constant. However, as highlighted in \cite{4a}, there are limitations to using the
$\upsilon_{s}^{2}$ as a measure of stability. Near $\omega = 1$, which corresponds to
the stiff matter region, Eq.\eqref{18} may not accurately represent the
$\upsilon_{s}^{2}$. This is because the microscopic properties of stiff matter are
not fully understood, so traditional fluid dynamics might not be applicable. Despite
these limitations, the $\upsilon_{s}^{2}$ condition provides a necessary, though not
sufficient, criterion for assessing the stability of the thin-shell around the
gravastar. Substituting the necessary values into \eqref{18}, we obtain
$\upsilon_{s}^{2}$. Figure \textbf{12} illustrates the behavior of $\upsilon_{s}^{2}$
with the radial distance $R$. The fact that the condition $ 0 \leq \upsilon_{s}^{2}
\leq 1$ is satisfied indicates that the gravastar maintains stability under these
conditions.
\begin{figure}\center
\epsfig{file=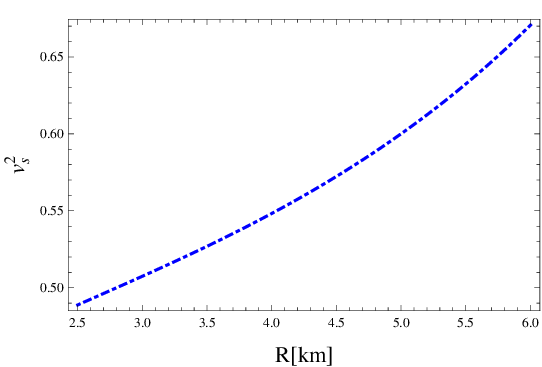,width=.48\linewidth} \caption{Plot of $\upsilon_{s}^{2}$ with
respect to $R$.}
\end{figure}
\begin{figure}\center
\epsfig{file=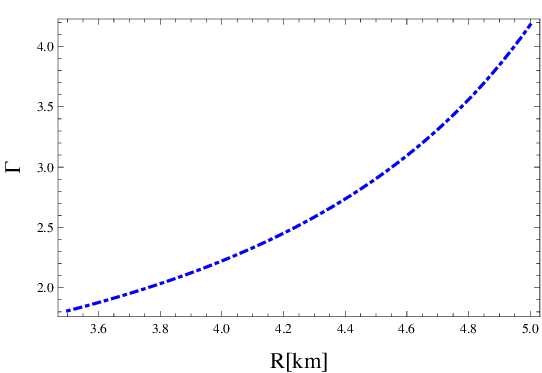,width=.47\linewidth} \caption{Plot of $\Gamma$ with respect to
$R$.}
\end{figure}

The adiabatic index $(\Gamma)$ is essential for evaluating the
stability of stars, such as gravastars. It represents the
correlation between pressure and density and how they affect a star
stability. This index is calculated as
\begin{equation}
\Gamma = \frac{\upsilon_{s}^{2} (p + \varrho)}{p}.
\end{equation}
The stability of an object is determined for $\Gamma$ to be greater than 4/3. If
$\Gamma$ is below this critical value, the object becomes unstable and could collapse
\cite{52a}. Figure \textbf{13} shows that the system meets the required stability
criterion, indicating that the gravastar is both viable and stable.

\section{Concluding Remarks}

Gravastars have been studied under various gravitational theories.
In this work, we have focused on exploring gravastars within the
extended STG framework, which is effective in describing the current
accelerated expansion of the universe. We have examined a gravastar
solution using the Finch-Skea metric to establish a model that is
both physically plausible and free of singularities, effectively
circumventing the issues associated with black hole like
singularities. By utilizing this metric with an ideal fluid
distribution, we have studied the properties of the gravastar
interior, shell and exterior regions. A summary of the results is
provided as follows.
\begin{itemize}
\item
The $p_r$ exceeds $p_t$ in the current model, leading to negative anisotropy and an
inward directed attractive force Figure \textbf{1}.
\item
We have observed that as the shell radius increases, its mass also
rises significantly, indicating a direct relationship between the
two. This increase in mass highlights the growing influence of the
shell as it expands, reflecting the model sensitivity to the
physical properties of the shell (Figure \textbf{2}).
\item
The pressure and density in the thin-shell increase towards the outer boundary, with
the outer edge becoming denser as the shell thickens. This difference in density and
pressure is important for understanding the strength of the gravastar, as it shows
that the outer shell is much more compressed as compared to the inner regions (Figure
\textbf{3}).
\item
We have observed that both the surface energy density and negative
pressure decrease as the shell thickness increases. The surface
energy density and pressure gradually diminish towards the outer
boundary. This decrease indicates that as the shell expands, the
energy stored per unit area is reduced, suggesting a more stable
outer region (Figure \textbf{4}).
\item
The energy inside rises as the shell thickness increases. This
suggests that the energy within the shell grows in direct proportion
to the radial distance. Additionally, the observed fluctuations
indicate that gravastars can sustain their structural stability
while steadily accumulating energy over time (Figure \textbf{5}).
\item
The proper length of the gravastar shell increases as its thickness grows, showing a
direct relationship between the two. This indicates that as the shell becomes
thicker, it expands in size, with the proper length reflecting the extent of this
growth (Figure \textbf{6}).
\item
We have ensured that the NEC is satisfied for a model parameter, indicating that the
gravastar shell could consist of ordinary matter. This suggests that gravastars can
exist as physically plausible objects (Figure \textbf{7}).
\item
For a stable gravastar, the entropy should reach its maximum value at the surface. We
have observed that the shell entropy increases as the shell thickness expands. Our
results are consistent with the previous studies \cite{4cc}-\cite{4d} (Figure
\textbf{8}).
\item
The EoS parameter drops below -1, entering the phantom region. This phantom-like
behavior further highlights the distinct characteristics of gravastars in this
modified gravity theory (Figure \textbf{9}).
\item
The effective potential shows that $V^{\prime\prime}(R) > 0$, meaning the thin-shell
is stable (Figure \textbf{10}).
\item
The surface redshift, causality condition and adiabatic index indicate that the
gravastar satisfies the stability criteria, hence the gravastar is stable (Figures
\textbf{11-13}).
\end{itemize}

In conclusion, our results present a physically plausible
alternative to black holes by avoiding singularities and event
horizons, although obtaining physically acceptable solutions with
this specific metric potential proved more challenging than in the
previous work \cite{11}. Our model provides a more accurate
analytical solution for determining the physical parameters of the
shell, thereby enhancing the understanding of gravastar stability.
This stability has been confirmed through different techniques. In
future, exploring more general forms of $f(\mathcal{Q}, \mathbb{T})$
gravity will be crucial for deepening our understanding of gravastar
solutions and assessing whether the same thickness-to-stability
relationships hold across different gravity models.

In this study, we have conducted a thorough examination of the
physical and stability characteristics of gravastars within the
$f(\mathcal{Q}, \mathbb{T})$ gravity framework, employing the
Finch-Skea metric to model their structure. By investigating
properties such as energy distribution, proper length, energy
conditions, entropy, and the EoS parameter, we have demonstrated the
gravastar remarkable adaptability and stability. Our findings
indicate that the energy inside the shell increases proportionally
with its thickness, highlighting its efficient energy accumulation
capability. The proper length also increases with thickness. The
null energy condition is satisfied in our model, confirming its
physical plausibility, while entropy increases with shell thickness.
These results collectively emphasize the gravastar potential as a
singularity-free alternative to black holes. Our stability analysis
has revealed that the effective potential exhibits a positive second
derivative, affirming the robustness of the thin shell under radial
perturbations. The surface redshift is below the critical limit of
2, while the causality condition $0 \leq v_s^2 \leq 1$ and the
adiabatic index $\Gamma > 4/3$ satisfy the key stability criteria.
Together, these findings confirm the gravastar stable configuration
within the $f(\mathcal{Q}, \mathbb{T})$ gravity framework.

Comparing with existing literature, our results align well with
Sharif and Saba \cite{a2} in $f(\mathcal{G}, \mathbb{T}^{2})$
gravity, where $\mathcal{G}$ is the Gauss-Bonnet invariant, using
the Finch-Skea metric with isotropic fluids, concluding their
stability and physical plausibility. Faisal et al. \cite{13a}
explored gravastar stability in $f(\mathcal{Q})$ gravity, finding
consistent structural stability under various conditions. Our
findings complement these studies, showcasing similar stability
criteria while extending the analysis to the $f(\mathcal{Q},
\mathbb{T})$ framework. By integrating the coupling between
non-metricity and the energy-momentum tensor trace, our work
provides new insights into gravastar properties, enhancing their
theoretical foundation and broadening the scope of modified gravity
theories. These results serve as a significant step forward in
understanding gravastars and their role as viable, singularity-free
alternatives to black holes.
\textbf{Data Availability Statement:} No new data were generated or
analyzed in support of this research.

\end{document}